\documentclass[aps,12pt,prc]{revtex4}
\usepackage{graphicx,epsf}

\begin{document}

\title{Nucleon Polarizabilities from Low-Energy Compton Scattering}

\author{{\bf S.R.~Beane}$^1$,
{\bf M.~Malheiro}$^{2}$,
{\bf J.A.~McGovern}$^{3}$,
{\bf D.R.~Phillips}$^{1,4}$, and
{\bf U.~van Kolck}$^{5,6}$ }

\vspace{20.0pt}

\affiliation{$^{1}$Department of Physics,
University of Washington, Seattle, WA 98195-1560, USA}

\affiliation{$^{2}$Instituto de F\'{\i}sica,  
Universidade Federal Fluminense, 24210-340, Niter\'oi, R.J., Brazil}

\affiliation{$^{3}$Department of Physics and Astronomy, 
University of Manchester, Manchester M13 9PL, UK}

\affiliation{$^{4}$Department of Physics and Astronomy,
Ohio University, Athens, OH 45701, USA}

\affiliation{$^{5}$Department of Physics,
University of Arizona, Tucson, AZ 85721, USA}

\affiliation{$^{6}$RIKEN BNL Research Center, 
Brookhaven National Laboratory, Upton, NY 11973, USA}

\begin{abstract} 
  An effective field theory is used to give a model-independent
  description of Compton scattering at energies comparable to the pion
  mass.  The amplitudes for scattering on the proton and the deuteron,
  calculated to fourth order in small momenta in chiral perturbation
  theory, contain four undetermined parameters that are in one-to-one
  correspondence with the nucleon polarizabilities.  These
  polarizabilities are extracted from fits to data on elastic photon
  scattering on hydrogen and deuterium. For the proton we find:
  $\alpha_p=(12.1 \pm 1.1~({\rm stat.}))_{-0.5}^{+0.5}~({\rm theory})$
  and $\beta_p=(3.4 \pm 1.1~({\rm stat.}))_{-0.1}^{+0.1}~({\rm
  theory})$, both in units of $10^{-4}~{\rm fm}^3$.  For the isoscalar
  polarizabilities we obtain: $\alpha_N=(13.0 \pm 1.9~({\rm
  stat.}))_{-1.5}^{+3.9}~({\rm theory})$ (in the same units) while $\beta_N$ is
  consistent with zero within sizable error bars. 
\end{abstract}

\maketitle

\vspace{0.8cm}

\newcommand{\lsc}{\Lambda _\chi}
\newcommand{\cpt}{$\chi$PT}
\newcommand{\sss}{\scriptscriptstyle }
\newcommand{\ga}{g_{\sss A}}
\newcommand{\yo}{F_\pi}
\newcommand{\vsigone}{{\vec\sigma^{\sss 1}}}
\newcommand{\vsigtwo}{{\vec\sigma^{\sss 2}}}
\newcommand{\veps}{{\vec\epsilon}}
\newcommand{\vepsprime}{{\vec\epsilon\, '}}
\newcommand{\vkay}{{\vec k}}
\newcommand{\vkayprime}{{{\vec k}\, '}}
\newcommand{\vpee}{{\vec p}}
\newcommand{\vpeeprime}{{{\vec p}\, '}}
\newcommand{\vsigma}{{\vec\sigma}}
\newcommand{\ompi}{{\frac{\omega}{m_\pi}}}
\newcommand{\tpi}{{\frac{t}{m_\pi^2}}}

Electromagnetic polarizabilities are a fundamental property of any
composite object. For example, atomic polarizabilities contain
information about the charge and current distributions that result
from the interactions of the protons, neutrons, and electrons inside
the atom.  Protons and neutrons are, in turn, complex objects composed
of quarks and gluons, with interactions governed by QCD.  It has long
been hoped that neutron and proton polarizabilities will give
important information about the strong-interaction dynamics of
QCD. For example, in a simple quark-model picture these
polarizabilities contain averaged information about the charge and
current distribution produced by the quarks inside the
nucleons~\cite{holstein}. In this paper we use an effective field
theory (EFT) of QCD to extract both proton and neutron
polarizabilities in a consistent and systematic manner from Compton
scattering data---the first EFT extraction of all these important
quantities within the same framework 
\cite{f***note}.
The background to this work will be found primarily
in two papers, Ref.~\cite{judith} for the proton, and
Ref.~\cite{therestofus} for the deuteron.  Further details of interest
to the specialist will be published elsewhere \cite{usagain}.

In an atomic or molecular system the polarizabilities are measured with
static fields.  Nuclear polarizabilities can analogously be determined
by the scattering of long-wavelength photons.  Experimental facilities
which accurately measure the energy of a photon beam using photon
tagging have made possible a new generation of experiments which probe
the low-energy structure of nucleons and nuclei.  In particular,
photon tagging can be used to measure Compton scattering on
weakly-bound systems, since it facilitates the separation of elastic
and inelastic cross sections.  At sufficiently low incoming (outgoing)
photon energy $\omega$ ($\omega'$) and momentum $\vec{k}$
($\vec{k}'$), the spin-averaged Compton scattering amplitude for any
nucleus is, in the nuclear rest frame:
\begin{equation}
T=\vepsprime \cdot \veps
\left(-\frac{{\cal Z}^2 e^2}{m_A} + 4 \pi \alpha \omega \omega'\right)
+ 4 \pi \beta \, \vepsprime\times \vec{k}' \cdot \veps\times \vec{k} 
+ \ldots,
\label{pionlessT}
\end{equation}
where $\veps$ and $\vepsprime$ are the polarization vectors of the
initial and final-state photons.
The first term in this
series is a consequence of gauge invariance, and is the Thomson limit
for low-energy scattering on a target of mass $m_A$ and charge
${\cal Z} e$. 
The coefficients of the second and third terms are 
the target electric and magnetic polarizabilities, $\alpha$ and
$\beta$, respectively. 
The polarizabilities can be separated by the angular dependence:
for example, at forward (backward) angles the amplitude
depends only on $\alpha +\beta$ ($\alpha -\beta$).
Other terms, represented by ``$\ldots$'', include higher powers
of energy and momentum and relativistic corrections \cite{definitive_DR}.

Hydrogen targets are used to determine proton polarizabilities
$\alpha_p$ and $\beta_p$ \cite{pdata}.  By contrast, the absence
of dense, stable, free neutron targets requires that the neutron
polarizabilities $\alpha_n$ and $\beta_n$ be extracted from scattering
on deuterium (or other nuclear) targets.  Data exist for coherent
$\gamma d\rightarrow \gamma d$ from 49 to 95 MeV
\cite{lucas,SAL,lund}, and for quasi-free 
$\gamma d\rightarrow \gamma pn$ from 200 to 400 MeV \cite{quasidata}.  The
coherent process is sensitive to the isoscalar nucleon 
polarizabilities---$\alpha_N \equiv (\alpha_p + \alpha_n)/2$, 
$\beta_N \equiv (\beta_p + \beta_n)/2$---via
interference with the larger Thomson term. The extraction of these
polarizabilities from data requires a consistent theoretical
framework that clearly separates nucleon properties from
nuclear effects. In the long-wavelength limit pertinent to
polarizabilities EFT provides a
model-independent way to do exactly this~\cite{kreview,bkmrev,bkreview}.

The EFT of QCD relevant to the low-energy interactions of a single
nucleon with any number of pions and photons is known as chiral
perturbation theory (\cpt). The  Lagrangian is constrained only by approximate 
chiral symmetry and the relevant space-time symmetries. 
$S$-matrix elements can be expressed as a simultaneous expansion in powers of
momenta and the pion mass (collectively denoted by $Q$)
over the characteristic scale of physics not
included explicitly in the EFT.  Many processes have been computed in this
EFT to nontrivial orders and it has proven remarkably successful
\cite{bkmrev}. In this, as in any EFT, detailed information about
short-distance physics is absent. The short-distance physics relevant
to low-energy processes appears in the theory as constants whose
determination lies outside the scope of the EFT itself.  In the
purest form of the theory they are determined by fitting experimental
observables. In many cases the sparsity of low-energy single-nucleon
data makes such a determination problematic---and for free neutrons
data is non-existent.

When the amplitude for unpolarized Compton scattering is expanded in
powers of $Q$, we obtain Eq.~(\ref{pionlessT}) with $\alpha$ and
$\beta$ given as functions of the EFT parameters.  To ${\cal O}
(Q^3)$, no parameters appear apart from those fit in other processes,
so predictions can be made.  At this order the proton and neutron
polarizabilities are given by pion-loop effects~\cite{ulf1}:
$\alpha_p=\alpha_n=10\beta_p=10\beta_n= 5 e^2 g_A^2/384 \pi^2 f_\pi^2
m_\pi =12.2 \times 10^{-4} \, {\rm fm}^3$. Here $g_A\simeq 1.26$ is
the axial coupling of the nucleon and $f_\pi\simeq 93$ MeV is the pion
decay constant.  At ${\cal O} (Q^4)$ there are new long-range
contributions to these polarizabilities. Four new parameters also
appear which encode contributions of short-distance physics to the 
spin-independent polarizabilities.  Thus,
minimally, one needs four pieces of experimental data to fix these
four short-distance contributions, but once they are fixed \cpt~makes
model-independent predictions for Compton scattering on protons and
neutrons.

The amplitude for Compton scattering on the nucleon has been computed
to ${\cal O} (Q^4)$ in Ref.~\cite{judith}. The shifts of $\alpha_p$
and $\beta_p$ from their $O(Q^3)$ values were not fitted directly to
the data in that work. Instead the Particle Data Group values for the
polarizabilities were used~\cite{PDG}. These were originally extracted
using a dispersion-theoretic approach that incorporates model-motivated
assumptions regarding the asymptotic behavior of the amplitude.  The
differential cross sections which result from this procedure are in
good agreement with the low-energy data~\cite{pdata}.

\begin{figure}[htb]
\epsfxsize=12.5cm
\centerline{\epsffile{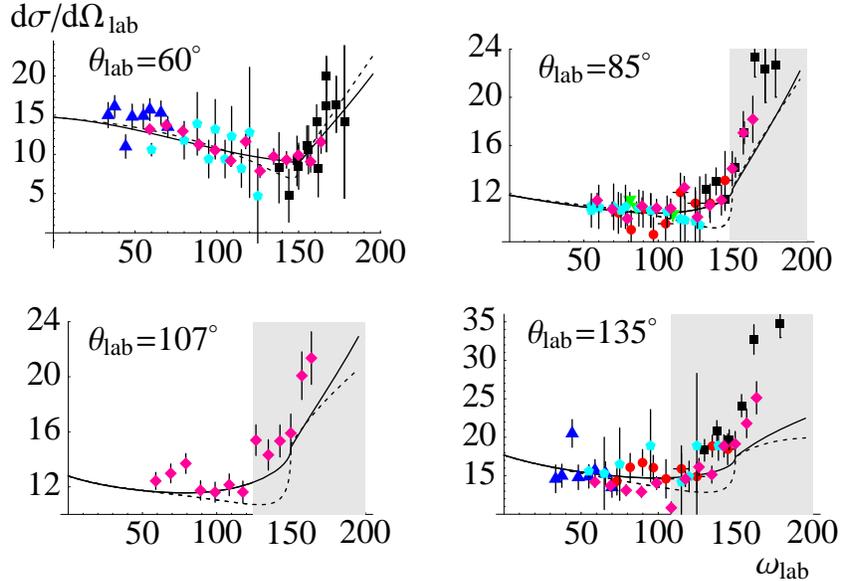}}
\caption {\label{Q4deuprotfit} Results of the ${\cal O}(Q^4)$ EFT best
fit (solid line) to the differential cross sections for Compton
scattering on the proton at four different angles, compared to the
experimental data \protect\cite{pdata,fitting}. (Symbols as in
Ref.~\protect\cite{judith}.) The grey area is the region excluded from
the fit ($\omega,\sqrt{|t|} > 200$ MeV). The dashed line is the 
${\cal O}(Q^3)$ prediction.  }
\end{figure}

Thus the calculation of Ref.~\cite{judith} is not, strictly
speaking, model independent: there is model input in the values
used for $\alpha_p$ and $\beta_p$. Using the same amplitude 
but fitting the polarizabilities, very good fits of the proton data
in the low-energy regime ($\omega, \sqrt{|t|}< 200$ MeV) can be obtained.
The central values for $\alpha_p$ and $\beta_p$ are similar to those
employed in Ref.~\cite{judith}, but the uncertainty is larger:
\begin{eqnarray}
\alpha_p &=& (12.1 \pm 1.1)_{-0.5}^{+0.5} \times 10^{-4} \, {\rm fm}^3, 
\nonumber\\
\beta_p &=& (3.4 \pm 1.1)_{-0.1}^{+0.1} \times 10^{-4} \, {\rm fm}^3, 
\label{eq:eval1}
\end{eqnarray}
where statistical (1-$\sigma$) errors are inside the brackets, and an
estimate of the contribution from higher-order terms is given
outside. These numbers are based on varying the upper bound on which
data are fit from 160 MeV to 200 MeV, and on estimates of the ${\cal
O}(Q^5)$ effect on $\gamma p$ scattering. A sample of the best-fit
results is shown, together with data, in Fig.~\ref{Q4deuprotfit}.  The
fit has $\chi^2/{\rm d.o.f.}=170/131$. These results are fully
compatible with other extractions, though the central value of
$\beta_p$ is somewhat higher~\cite{PDG,drechsel}. 
They are also consistent with
the values predicted in Ref.~\cite{ulf1,bernard},
where resonance saturation was used to estimate
the ${\cal O}(Q^4)$ short-distance
contributions. 

\begin{figure}[htb]
\epsfxsize=0.5\columnwidth
\centerline{\epsffile{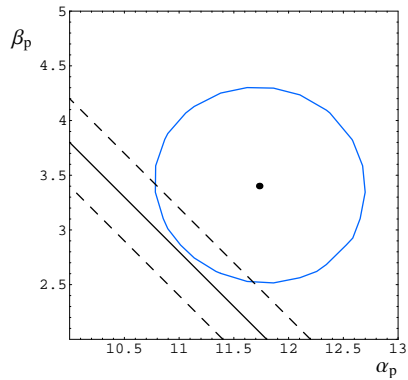}}
\caption {\label{fig-Baldincomp} The 1-$\sigma$ error ellipse for our fit
to the proton data with $\omega,\sqrt{|t|} < 200$ MeV. Also shown is the
band defined by the Baldin Sum Rule result (\ref{eq:Baldin}). The units
of $\alpha_p$ and $\beta_p$ are $10^{-4}~{\rm fm}^3$.}
\end{figure}

The values (\ref{eq:eval1}) can be compared with the recent
re-evaluation of the Baldin Sum Rule by Olmos de Le\'on {\it et
al.}~\cite{pdata}, which gives:
\begin{equation}
\alpha_p + \beta_p=13.8 \pm 0.4.
\label{eq:Baldin}
\end{equation}
This result overlaps the 1-$\sigma$ error ellipse of the fit
(\ref{eq:eval1}), as shown in Fig.~\ref{fig-Baldincomp}.  Including
the constraint (\ref{eq:Baldin}) in our fit leads to values for
$\alpha_p$ and $\beta_p$ consistent with (\ref{eq:eval1}), but with
smaller statistical errors, namely:
\begin{eqnarray}
\alpha_p &=& (11.0 \pm 0.5 \pm 0.2)_{-0.5}^{+0.5} \times 10^{-4} \, {\rm fm}^3,
\nonumber\\
\beta_p &=& (2.8 \pm 0.5 \mp 0.2)_{-0.1}^{+0.1} \times 10^{-4} \, {\rm fm}^3.
\label{eq:eval2}
\end{eqnarray}
In (\ref{eq:eval2}) we have left the systematic error unchanged, but
have now included a second error inside the brackets, whose source is
the error on the sum rule evaluation (\ref{eq:Baldin}). The smaller
overall statistical error in (\ref{eq:eval2}) is achieved at the
expense of additional (reasonable) assumptions about the high-energy
behaviour of hadronic amplitudes, and also by using higher-energy data
on the total proton photo-absorption cross section, since these are
the two ingredients entering the Baldin Sum Rule result for $\alpha_p
+ \beta_p$.

The amplitude for Compton scattering on a nuclear target can also be
calculated in the EFT, although the unraveling of scales is more
subtle when more than one nucleon is present \cite{bkreview}.  In the
single-nucleon sector a typical intermediate state has an energy
denominator of ${\cal O} (Q)$, but in multi-nucleon processes
``reducible'' intermediate states can have small energy denominators
of ${\cal O} (Q^2/m_N)$.  The resulting infrared enhancement
complicates the perturbative expansion~\cite{weinberg}.  Furthermore,
the very existence of nuclei implies a breakdown of perturbation theory;
hence the leading two-nucleon interactions should be summed
to all orders, thereby building up a nuclear wave function
$|\psi\rangle$.  In the case of Compton scattering at energies of
${\cal O} (m_\pi)$, this is the only resummation necessary
\cite{therestofus}, and the full amplitude, $T$, can be written in terms of
a kernel $K_{\gamma\gamma}$ which contains all irreducible (in the
above sense) $\gamma NN \rightarrow \gamma NN$ graphs:
\begin{equation}
T = \langle\psi|K_{\gamma\gamma}|\psi\rangle.
\label{nuclearT}
\end{equation}
The kernel can be calculated in \cpt. Several other reactions involving
deuterium have been successfully analyzed using analogous
approaches~\cite{bkreview}. However, for Compton scattering at energies of
${\cal O} (m_\pi^2/m_N)$ further resummations are necessary.  In
particular, such resummations restore the Thomson limit for Compton
scattering from deuterium~\cite{FT,springer,therestofus}. In 
this energy regime it seems more appropriate to use a lower-energy EFT
where pions have been integrated out~\cite{BS,GR00}.

At momentum transfers of ${\cal O} (m_\pi)$, a formally-consistent
power counting is emerging~\cite{chiralexp} which organizes the
nuclear interactions that give rise to the wave function $|\psi
\rangle$.  This power counting is an improvement over Weinberg's
original proposal \cite{weinberg}, which has led to fairly accurate
wave functions~\cite{bkreview}. To the order we are working in the
EFT, we require a deuteron wave function calculated to ${\cal O}(Q^2)$
in the chiral expansion. We employ a wave function generated using the
${\cal O}(Q^2)$ $\chi$PT potential of Ref.~\cite{chipots}. (The
$\Lambda=600$ MeV wave function was chosen, but choosing $\Lambda=540$
MeV instead produces very similar results.)  In order to test the
consistency of our error estimates, we have also computed results with
other ``realistic'' wave functions, such as that obtained from the
Nijm93 OBE potential~\cite{Nijm93}. Below we will quote results only
for the ${\cal O}(Q^2)$ wave function of Ref.~\cite{chipots} and for
the Nijm93 wave function. The results for differential cross sections
generated using $|\psi \rangle$'s obtained from other $NN$ potentials
almost invariably lie between these two extremes. 

Meanwhile, the kernel $K_{\gamma \gamma}$ is the sum of the
single-nucleon Compton amplitude, in which one nucleon is a
``spectator'' to the Compton scattering, and ``two-nucleon''
contributions in which both nucleons are involved in the scattering of
the photon.  The amplitude for coherent Compton scattering on the
deuteron was computed to ${\cal O} (Q^3)$ in Ref.~\cite{therestofus}.
There are no free parameters to this order. The corresponding cross
section is in good agreement with the Illinois data~\cite{lucas} at 49
and 69 MeV, but under-predicts the SAL data~\cite{SAL} at 95 MeV. This
calculation yields cross sections which agree well with the recent
Lund data~\cite{lund}. It also agrees qualitatively with
potential-model calculations of $\gamma d$ scattering~\cite{potmods,LL}.

We have now extended our calculation to ${\cal O} (Q^4)$. The
calculation contains both single-nucleon and two-nucleon
contributions. For the first class of diagrams, we employ the
single-nucleon amplitude of Ref.~\cite{judith}. The amplitude given
there must be boosted from the Breit frame to the $\gamma d$
center-of-mass frame, which is straightforward once the $\gamma p$
amplitude has been decomposed into six invariant functions
multiplying structures that transform in well-defined ways under
boosts~\cite{therestofus,usagain}. 

Less straightforward are technical issues associated with the fashion
in which nucleon recoil is included in the $\gamma d$ scattering
calculation. These occur because the heavy-baryon formulation of
$\chi$PT employed in this work expands about the limit of static
nucleons, with nucleon recoil treated as a perturbative
correction. 
The first issue has to do with the treatment of the very-low-energy
region, $\omega \sim m_\pi^2/M$. At these energies the nucleons inside
the deuteron are not static on the time-scales relevant to Compton
scattering, and so a resummation of nucleon-recoil corrections must be
performed. Below we will invoke this resummation when we calculate
$\gamma d$ scattering for these low photon energies.  The second issue
has to do with ensuring that particle-production cuts appear at the
correct position.  In the absence of nucleon recoil the channel
$\gamma d \rightarrow \pi d$ will open at a position which is in error
by an amount $m_\pi^2/(2 M_d)$. Since the opening of this channel can
result in rapid variation of the Compton cross section with energy,
this problem must be dealt with when we describe $\gamma p$ scattering
for $\omega$ close to $m_\pi$~\cite{bernard,judith}. However, even at the
highest energy for which Compton scattering on deuterium is calculated
below, this small error in the position of the $\pi d$ cut is a
relatively minor effect. We emphasize that these subtleties are
artifacts which arise due to the fact that nucleon recoil is treated
as a perturbation in the $\gamma$N amplitude we have employed
here~\cite{judith}. Such a treatment is in accord with the usual
practice in the chiral EFT known as heavy-baryon $\chi$PT, but a
perturbative treatment of nucleon recoil in not mandatory in the EFT.
Thus these two problems in no way represent true limitations of a
chiral EFT of Compton scattering.

In addition, the two-nucleon diagrams at ${\cal O} (Q^4)$ are added to
the two-nucleon ${\cal O}(Q^3)$ diagrams shown in
Ref.~\cite{therestofus}.  These ${\cal O}(Q^3)$ graphs were generated
by the leading chiral Lagrangian. In going to ${\cal O}(Q^4)$ we
include diagrams with one insertion from the sub-leading chiral
Lagrangian. The coefficients of these vertices are essentially
determined by relativistic invariance, and so these ${\cal O}(Q^4)$
two-body effects are suppressed by $Q/m_N$. No unknown parameters
occur in these graphs.  However, two free parameters associated with
the neutron polarizabilities do appear in the single-nucleon
contribution: the shifts in the isoscalar polarizabilities from their
${\cal O}(Q^3)$ values. Indeed, of all the additional graphs which
enter our new, ${\cal O}(Q^4)$, calculation only this effect from
single-nucleon Compton scattering changes the cross section
significantly.  (Details can be found in Ref.~\cite{usagain}.)

We have fitted these two free parameters to the existing $\gamma d$
scattering data. As in the proton case, we impose a cutoff on the data
that we fit when we extract $\alpha$ and $\beta$. With a cutoff of
$\omega,\sqrt{|t|} < 200$~MeV all of the 29 world data
points~\cite{lucas,SAL,lund} are included. We also use a cutoff of 160
MeV, in which case all but the two backward-angle SAL points must be
fitted. In both cases we float the experimental normalization for each
experimental run within the quoted systematic error~\cite{fitting},
resulting in 22 (20) degrees of freedom for the fit~\cite{usagain}.

Using the wave function of Ref.~\cite{chipots}, we fit to data with
$\omega, \sqrt{|t|} < 160$ MeV. This produces the isoscalar nucleon
polarizabilities and the $\chi^2$ per degree of freedom given in the
first line of Table~\ref{table-results}. The $\chi^2$ is unacceptably
large. This is driven mainly by the 49 MeV data from Illinois, and
occurs because, as discussed above, for $\omega \sim m_\pi^2/m_N$
further resummations are necessary in the EFT.  Here we adopt the
strategy of including the dominant contributions that need to be added
to our standard ${\cal O}(Q^4)$ result in this very-low-energy region
in order to restore the Thomson limit for the $\gamma d$
amplitude~\cite{therestofus}. This yields the results on the second
line of Table~\ref{table-results}. An alternative strategy, namely
dropping the 49 MeV data altogether, produces very similar central
values for $\alpha_N$ and $\beta_N$.

\begin{table*}[htb]
\begin{ruledtabular}
\begin{tabular}{|c|c|c|c||c|c|c|}
Chiral & Wave & Very-low-energy 
& $\omega, \sqrt{|t|} $ &
$\alpha_N$ & $\beta_N$ & $\chi^2$/d.o.f. \\
order & function & resummation? & & $(10^{-4}~{\rm fm}^3)$ & $(10^{-4}~{\rm fm}^3)$&\\
\hline
$Q^4$ & NLO $\chi$PT & No  & $200$ MeV & 13.6 & $0.1$ & 2.36 \\
$Q^4$ & NLO $\chi$PT & No  & $160$ MeV & 15.4 & $-2.3$ & 1.95 \\
$Q^4$ & NLO $\chi$PT & Yes & $160$ MeV & 13.0  & $-1.8$ & 1.33\\
$Q^4$ & Nijm93       & Yes & $200$ MeV & 16.9 & $-2.7$ & 2.87 \\\end{tabular}
\end{ruledtabular}
\caption{Results from different $\chi$PT extractions of isoscalar nucleon
polarizabilities from $\gamma d$ scattering data.}
\label{table-results}
\end{table*}

When the upper limit on which data is fitted is increased the central
value of $\alpha_N-\beta_N$ changes markedly, with 
a slightly higher $\chi^2$/d.o.f. (third line of Table~\ref{table-results}). 
Another key test involves examining the
impact of the choice of deuteron wave function. The
variability in the differential cross section
due to the choice of wave function is $\sim 10$\%.
When the Nijm93 wave function is used the
results are as shown in the fourth line of
Table~\ref{table-results}. The high $\chi^2$ is again due to a failure
to reproduce the 49 MeV data---this time even {\it with} the
very-low-energy contributions included. While the Nijm93 wave function
is not consistent with $\chi$PT, it does have the correct
long-distance behavior. Such sensitivity to the choice of $|\psi
\rangle$ is worrisome and merits further study.

\begin{figure}[htb]
\vspace{0.2in} 
\epsfxsize=11.0cm
\centerline{\epsffile{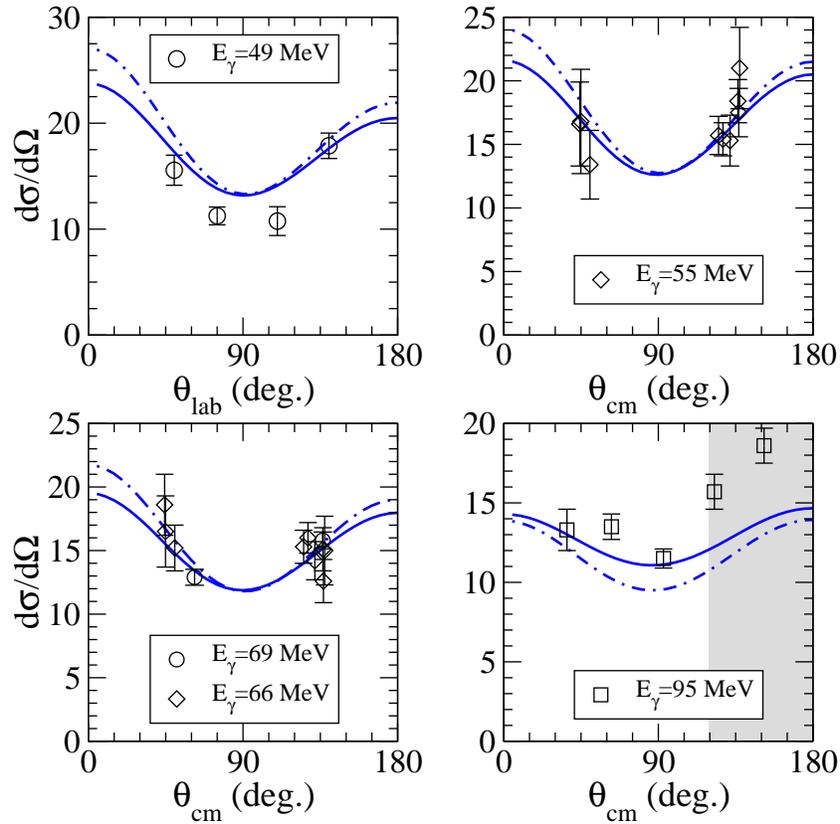}}
\caption {\label{Q4deutfit} Results of the $O(Q^4)$ EFT best fit to
the (lab and c.m. as appropriate) differential cross sections for
Compton scattering on deuterium at four different lab photon energies:
49, 55, 67, and 94.5 MeV. The data are from Illinois~\protect\cite{lucas}
(circles), Lund~\protect\cite{lund} (diamonds) and SAL
(squares)~\protect\cite{SAL}. The error bars represent the quoted statistical
(only) uncertainties of these measurements. The solid line is the
$O(Q^4)$ calculation with $\alpha_N=13.0 \times 10^{-4}~{\rm fm}^3$,
$\beta_N=-1.8 \times 10^{-4}~{\rm fm}^3$.  The gray area is
the region excluded from the fit ($\omega,\sqrt{|t|} > 160$ MeV).  The
dot-dashed line is the (parameter-free) $O(Q^3)$ calculation.}
\end{figure}

Putting these results together we conclude that our best fit, and error bars,
for the {\it isoscalar  nucleon} polarizabilities are:
\begin{eqnarray}
\alpha_N &=& (13.0 \pm 1.9)_{-1.5}^{+3.9} \times 10^{-4} \, {\rm fm}^3,
\nonumber\\
\beta_N &=& (-1.8 \pm 1.9)_{-0.9}^{+2.1} \times 10^{-4} \,
{\rm fm}^3.
\label{eq:Npols}
\end{eqnarray}
The statistical errors inside the brackets are obtained from the
boundary of the 68\% C.L. region. The errors outside the brackets
reflect the arbitrariness as to which data are included, and which
deuteron wave function is employed.  The results for the best-fit EFT
are shown in Fig.~\ref{Q4deutfit}.

The results (\ref{eq:Npols}) differ from those given in the published
version of this work~\cite{allofus}, since after publication we
discovered two errors in our analysis of the $\gamma$d
data~\cite{lucas,SAL,lund}.  First, we mistakenly fitted the SAL data
at a c.m. energy of 95 MeV, whereas we should have employed a lab
energy of 94.5 MeV~\cite{SAL}.  (A similar error that ultimately has a
much smaller impact was made with the data from Lund~\cite{lund}.)
This causes a relatively small change in the results for most of our
fits, but it does increase the central value for $\alpha_N$ obtained
with the Nijm93 deuteron wave function by $1.0 \times 10^{-4}~{\rm
fm}^3$ over the value quoted in the published version of
Ref.~\cite{allofus}~\footnote{Thanks to Robert Hildebrandt and Jerry
Feldman for pointing out this error to us.}. A larger effect occurs
because there was an error in the program used to generate the
$O(Q^4)$ results presented for $\gamma$d scattering in
Ref.~\cite{allofus}.  This affects the $O(Q^3)$ results by less than
1\%. However, at $O(Q^4)$ the omission of this factor modifies the
interference between the spin-dependent pieces of the single-nucleon
amplitude and the $O(Q^4)$ two-body currents~\footnote{Thanks to
Deepshikha Choudhury for pointing out this mistake to us.}.
Correcting this mistake increases the predicted cross sections by
about 10-15\% with respect to those published in Ref.~\cite{allofus}.
We emphasize that neither of these mistakes affects any of the results
for $\gamma$p scattering reported here and in the published version of
Ref.~\cite{allofus}.  For more details see Ref.~\cite{usagain}.

Combining the revised numbers (\ref{eq:Npols}) with our results for
$\alpha_p$ and $\beta_p$ we see that a wide range of neutron
polarizabilities is consistent with a model-independent analysis of
the current low-energy proton and coherent deuteron Compton data.
Narrower ranges for the neutron polarizabilities can be obtained from
this data at the expense of introducing model dependence.  The values
of $\alpha_n$ and $\beta_n$ extracted from data on the quasi-free
process $\gamma d \rightarrow \gamma n p$ using a theoretical model
fall within this range~\cite{quasidata}.  However, there is clear
statistical evidence in these data sets for $\alpha_n$ being
significantly larger than $\beta_n$: $\beta_n$ is consistent with zero
within our large error bars. Also, the results presented here provide
{\it no} evidence for significant isovector components in $\alpha$ and
$\beta$.
 
In conclusion, we have determined nucleon polarizabilities from a
model-independent fit to low-energy Compton scattering on the proton
and the deuteron. Our results are consistent, within error bars, with
the recent extraction of $\alpha_N \pm \beta_N$ from the Lund data
using the detailed model of Levchuk and L'vov~\cite{lund,LL}. (But see
also the values found using the data of Refs.~\cite{lucas,SAL} and
this model~\cite{SAL,LL}.).  They are also consistent with the Baldin
sum rule results for $\alpha_p + \beta_p$ and $\alpha_n +
\beta_n$~\cite{baldin}.  The EFT can be improved by the introduction
of an explicit $\Delta$-isobar field, a complete treatment of the
very-low-energy region, and a better understanding of the dependence
of the results on the choice of deuteron wave function. An EFT study
of the quasi-free deuteron process is also an important future step.

\begin{acknowledgments}
We thank T. Hemmert and M.~Lucas for discussions, E.~Epelbaum and
V.~Stoks for providing us with deuteron wave functions, and J.~Brower
for coding assistance.  MM, DRP, and UvK thank the Nuclear Theory
Group at the University of Washington for hospitality while part of
this work was carried out, and UvK thanks RIKEN, Brookhaven National
Laboratory and the US DOE [DE-AC02-98CH10886] for providing the
facilities essential for the completion of this work.  This research
is supported in part by the US DOE under grants DE-FG03-97ER41014
(SRB), DE-FG02-93ER40756 (DRP), by the UK EPSRC
(JM), by Brazil's CNPq (MM), by DOE OJI Awards (DRP,UvK) and by an
Alfred P. Sloan Fellowship (UvK).
\end{acknowledgments}

\end{document}